\documentclass[journal]{IEEEtran}
\usepackage{amsfonts}
\usepackage{amsmath}
\usepackage{amsthm}
\usepackage{amssymb}
\usepackage{graphicx}
\usepackage[T1]{fontenc}
\usepackage{supertabular}
\usepackage{longtable}
\usepackage[usenames,dvipsnames]{color}
\usepackage{bbm}
\usepackage{fancyhdr}
\usepackage{breqn}
\usepackage{fixltx2e}
\usepackage{capt-of}
\usepackage[framemethod=TikZ]{mdframed}
\usepackage{xcolor}
\usepackage{tikz}
\usepackage{endnotes}
\usepackage{supertabular}
\usepackage{tabularx}
\usepackage{lipsum}
\usepackage{soul}

\newtheorem{proposition}{Proposition}
\newtheorem{definition}{Definition}

\DeclareMathOperator{\sgn}{sgn}
\newcommand{\mathsym}[1]{{}}
\newcommand{\unicode}[1]{{}}

\usepackage{grffile}
\mdfsetup{ middlelinecolor=black,
   middlelinewidth=1pt,
   backgroundcolor=yellow!40,
   roundcorner=5pt}

\begin{document}

\title{\color{Brown} How Much Data Do You Need? A Pre-asymptotic Metric for Fat-tailedness}
\author{Nassim Nicholas Taleb\\
Tandon School of Engineering, New York University\\
November 2018\\
Forthcoming, \textit{International Journal of Forecasting}\\

\thanks{ 
\color{Brown} The author owes the most to the focused comments by Michail Loulakis who, in addition, provided the rigorous derivations for the limits of the $\kappa$ for the Student T and lognormal distributions, as well as to the patience and wisdom of Spyros Makridakis. The paper was initially presented at \textit{Extremes and Risks in Higher Dimensions}, Sept 12-16 2016, at the Lorentz Center, Leiden and at Jim Gatheral's Festschrift at the Courant Institute, in October 2017. The author thanks Jean-Philippe Bouchaud, John Einmahl, Pasquale Cirillo, and others. Laurens de Haan suggested changing the name of the metric from "gamma" to "kappa" to avoid confusion. Additional thanks to  Colman Humphrey, Michael Lawler, Daniel Dufresne and others for discussions and insights with derivations.

}

}

\maketitle
\thispagestyle{fancy}
\markboth{\textbf{FAT TAILS STATISTICAL PROJECT}}
\flushbottom 
\begin{abstract}
\begin{mdframed}

This paper presents an operational metric for univariate unimodal probability distributions with finite first moment, in $[0,1]$ where 0 is maximally thin-tailed (Gaussian) and 1 is maximally fat-tailed. It is based on "how much data one needs to make meaningful statements about a given dataset?"

\underline{Applications}: Among others, it \begin{itemize}
 \item	helps assess the sample size $n$ needed for statistical significance outside the Gaussian,
  \item	helps measure the speed of convergence to the Gaussian (or stable basin),
 \item allows practical comparisons across classes of fat-tailed distributions,
 \item  allows the assessment of the number of securities needed in portfolio construction to achieve a certain level of risk-reduction from diversification,
 \item helps assess risks under various settings,
\item helps understand some inconsistent attributes of the lognormal, pending on the parametrization of its variance. 
 \end{itemize}
 
The literature is rich for what concerns asymptotic behavior, but there is a large void for finite values of $n$, those needed for operational purposes.

\underline{Background}: Conventional measures of fat-tailedness, namely 1) the tail index for the power law class, and 2) Kurtosis for finite moment distributions fail to apply to some distributions, and do not allow comparisons across classes and parametrization, that is between power laws outside the Levy-Stable basin, or power laws to distributions in other classes, or power laws for different number of summands. How can one compare a sum of 100 Student T distributed random variables with 3 degrees of freedom to one in a Levy-Stable or a Lognormal class? How can one compare a sum of 100 Student T with 3 degrees of freedom to a single Student T with 2 degrees of freedom?
 
We propose an operational and heuristic metric that allows us to compare $n$-summed independent variables under all distributions with finite first moment. The method is based on the rate of convergence of the law of large numbers for finite sums, $n$-summands specifically.
 
We get either explicit expressions or simulation results and bounds for the lognormal, exponential, Pareto, and the Student T distributions in their various calibrations --in addition to the general Pearson classes.

\end{mdframed}

%
%
%

\end{abstract}

 \begin{figure}[h!]
\includegraphics[width=\linewidth]{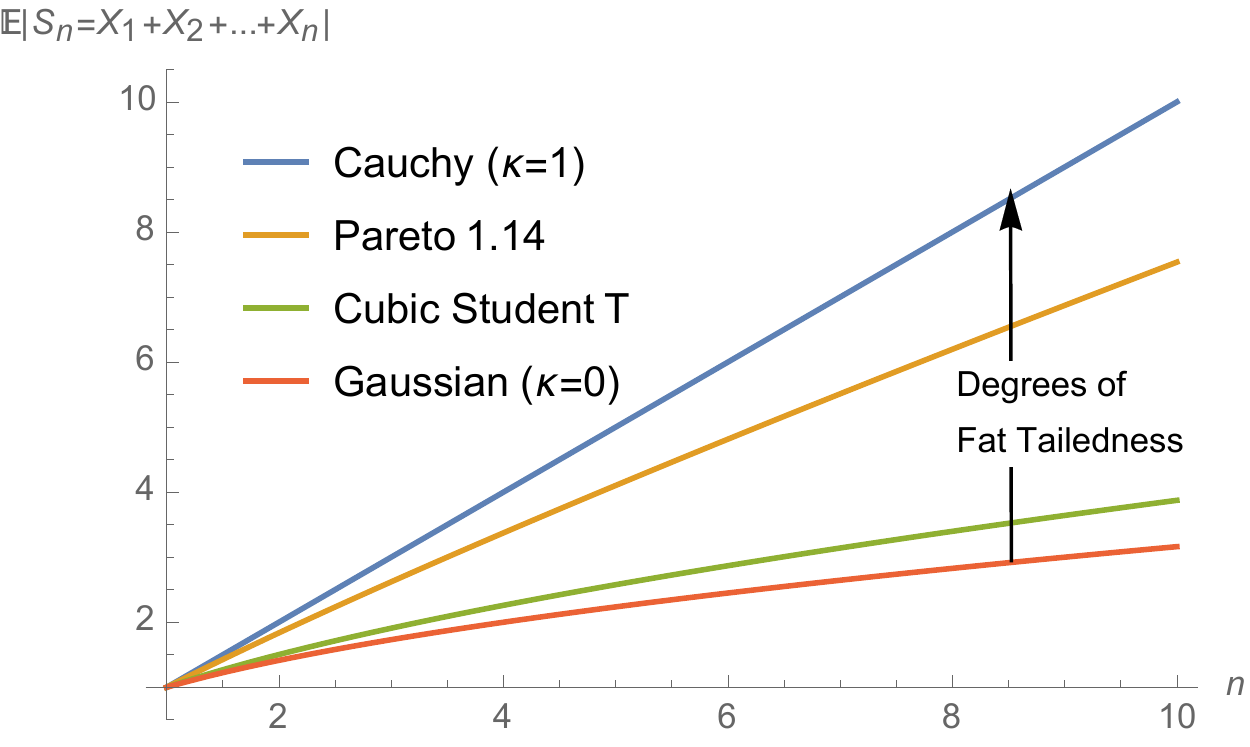}
\caption{The intuition of what $\kappa$ is measuring: how the mean deviation of the sum of identical copies of a r.v. $S_n= X_1+X_2+\ldots X_n$ grows as the sample increases and how we can compare preasymptotically distributions from different classes.} 
\end{figure}

 \begin{figure}[h!]
\includegraphics[width=\linewidth]{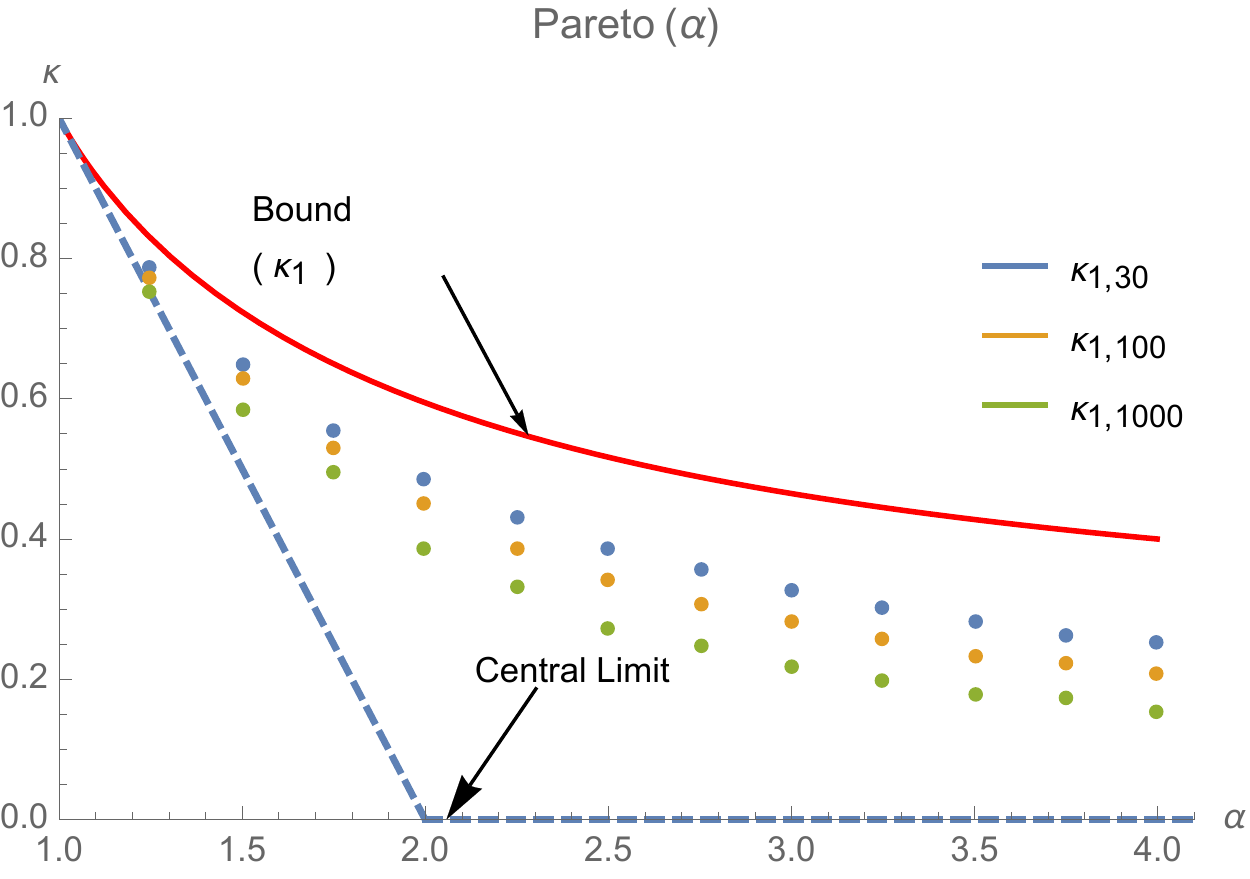}
\includegraphics[width=\linewidth]{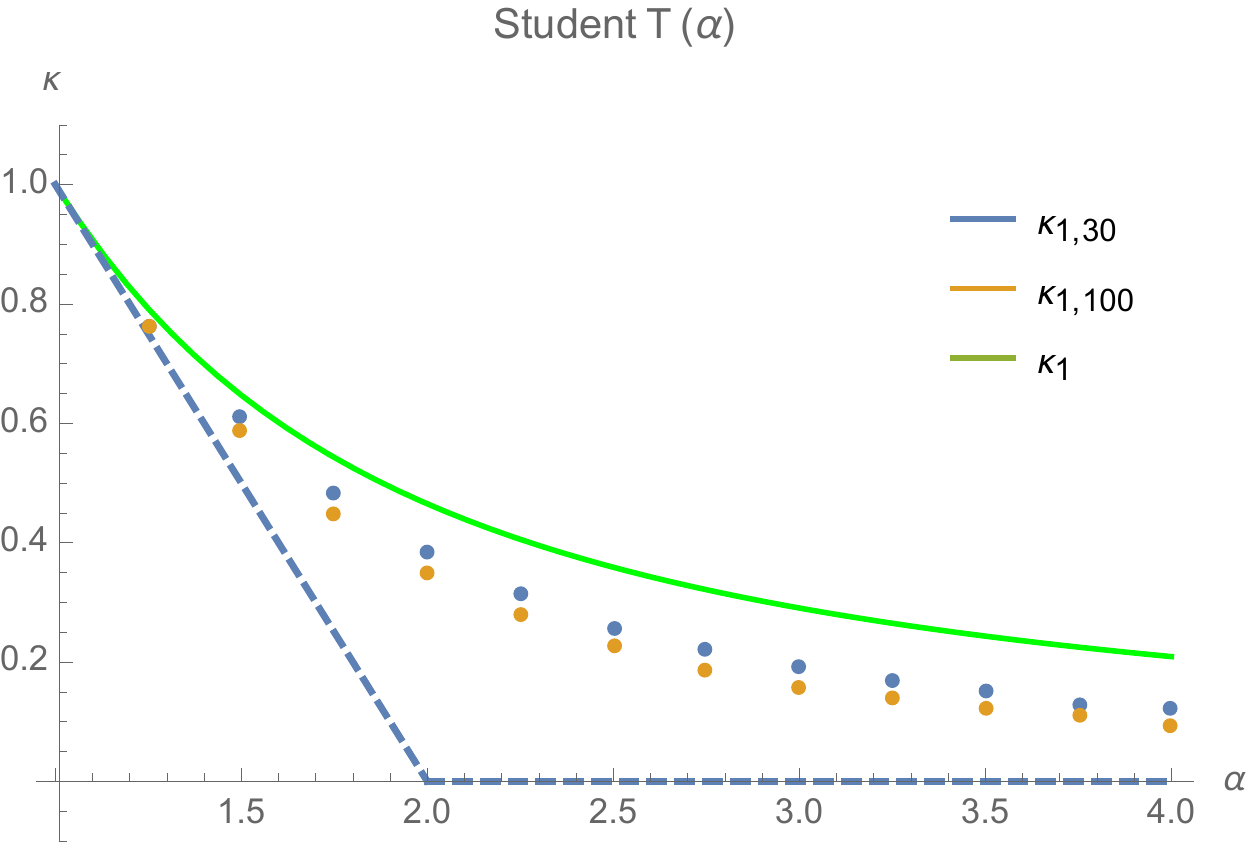}
\caption{Watching the effect of the Generalized Central Limit Theorem: Pareto and Student T Distribution, in the $\mathfrak{P}$ class, with $\alpha$ exponent, $\kappa$ converge to $2-(\mathbbm{1}_{\alpha<2} \alpha + \mathbbm{1}_{\alpha\geq 2} 2)$, or the Stable $\mathfrak{S}$ class. We observe how slow the convergence, even after 1000 summands. This discounts Mandelbrot's assertion that an infinite variance Pareto can be  subsumed into a stable distribution.} \label{convergencetoCLT}
\end{figure}

\section{Introduction and Definitions}

How can one compare a Pareto distribution with tail $\alpha=2.1$ that is, with finite variance, to a Gaussian? Asymptotically, these distributions in the regular variation class with finite second moment, under summation, become Gaussian, but pre-asymptotically, we have no standard way of comparing them given that metrics that depend on higher moments, such as kurtosis, cannot be of help. Nor can we easily compare an infinite variance Pareto distribution to its limiting $\alpha$-Stable distribution (when both have the same tail index or tail exponent). Likewise, how can one compare the "fat-tailedness" of, say a Student T with $3$ degrees of freedom to that of a Levy-Stable with tail exponent of $1.95$? Both distributions have a finite mean; of the two, only the first has a finite variance but, for a small number of summands, behaves more "fat-tailed" according to some operational criteria.\footnote{ By "fat tails" we are using the generic term used by finance practitioners to refer to thicker tails than the Gaussian, without reference to any particular class of distributions. }

\subsubsection{Criterion for "fat-tailedness"} There are various ways to "define" Fat Tails and rank distributions according to each definition. In the narrow class of distributions having all moments finite, it is the kurtosis, which allows simple comparisons and measure departures from the Gaussian, which is used as a norm. For the power law class, it can be the tail exponent. One can also use extremal values, taking the probability of exceeding a maximum value, adjusted by the scale (as practiced in extreme value theory). For operational uses, practitioners' fat-tailedness is a degree of concentration, such as "how much of the statistical properties will be attributable to a single observation?", or, appropriately adjusted by the scale (or the mean dispersion), "how much is the total wealth of a country in the hands of the richest individual?" 

Here we use the following criterion for our purpose, which maps to the measure of concentration in the past paragraph: "How much will additional data (under such a probability distribution) help increase the stability of the observed mean". The purpose is not entirely statistical: it can equally mean: "How much will adding an additional security into my portfolio allocation (i.e., keeping the total constant) increase its stability?"

Our metric differs from the asymptotic measures (particularly ones used in extreme value theory) in the fact that it is fundamentally preasymptotic.

Real life, and real world realizations, are outside the asymptote.

\subsubsection{What does the metric do}
The metric we propose, $\kappa$ does the following:
\begin{itemize}
\item Allows  comparison of $n-$summed variables of different distributions for a given number of summands , or same distribution for different $n$, and assess the preasymptotic properties of a given distributions.
\item Provides a measure of the distance from the limiting distribution, namely the L\'evy $\alpha$-Stable basin (of which the Gaussian is a special case).
\item For statistical inference, allows assessing the "speed" of the law of large numbers, expressed in change of the mean absolute error around the average thanks to the increase of sample size $n$.
 \item Allows  comparative assessment of the "fat-tailedness" of two different univariate distributions, when both have finite first moment.
 \item Allows us to know ahead of time how many runs we need for a Monte Carlo simulation.
\end{itemize}

\subsubsection{The state of statistical inference}
The last point, the "speed", appears to have been ignored. For in the 9,400 pages of the \textit{Encyclopedia of Statistical Science} \cite{encyclopediastat2004}, we were unable to find a single comment as to how long it takes to reach the asymptote, or how to deal with $n$ summands that are large but perhaps not sufficiently so for the so-called "normal approximation".
Further, the entry on statistical inference (authored by W. Hoeffding) explicitly brushes away the problem, stating:
\begin{quotation}
	 "The exact distribution of a statistic is usually highly complicated and difficult to work with. Hence the need to approximate the exact distribution by a distribution of a simpler form whose properties are more transparent. The limit theorems of probability theory provide an important tool for such approximations. In particular, the classical central limit theorems state that the sum of a large number of independent random variables is approximately normally distributed under general conditions. In fact, the normal distribution plays a dominating role among the possible limit distributions.
	 
 (...)	 Moreover, many statistics behave asymptotically like sums of independent random variables. All of this helps to explain the importance of the normal distribution as an asymptotic distribution."
	 
\end{quotation}
Even social science discussions of the "law of small numbers" \cite{tversky1971belief} assume Gaussian attributes as the norm. As to extreme value theory, the "functional law of small numbers" \cite{falk2010laws} concerns Poisson hitting with small probabilities; more generally, extreme value theory (while naturally equipped with the tools for fat tails) is concerned with the behavior of maxima, not averages.

Our motto here and elsewere is "statistics is never standard". This metric aims at showing \textit{how standard is standard}, and measure the exact departure \textit{from the standard} from the standpoint of statistical significance.

\section{The Metric}

	\begin{table*}[h]
\caption{Kappa for 2 summands, $\kappa_1$.}	
\begin{mdframed}

\begin{tabular}{p{0.15\linewidth}|c}

  \textbf{Distribution} & $\kappa_1$ \\
  
\hline
 & \\
 Student T $(\alpha)$  & $2-\frac{2\log (2)}{2 \log \left(\frac{2^{2-\alpha } \Gamma \left(\alpha -\frac{1}{2}\right)}{\Gamma \left(\frac{\alpha }{2}\right)^2}\right)+\log (\pi )}$\\
 &\\
 Exponential/Gamma & $2-\frac{\log (2)}{2 \log (2)-1} \approx .21$ \\
  & \\
 Pareto $(\alpha)$ & $2-\frac{\log (2)}{\log \left((\alpha -1)^{2-\alpha } \alpha ^{\alpha -1} \int_0^{\frac{2}{\alpha -1}} -2 \alpha ^2 (y+2)^{-2 \alpha -1} \left(\frac{2}{\alpha -1}-y\right) \left(B_{\frac{1}{y+2}}(-\alpha ,1-\alpha )-B_{\frac{y+1}{y+2}}(-\alpha ,1-\alpha )\right) \, dy\right)}$\footnote{$B_.(.,.)$ is the incomplete Beta function: $B_z(a,b)=\int _0^z t^{a-1} (1-t)^{b-1} dt$; $\text{erf}(.)$ is the error function $\text{erf}(z)=\frac{2}{\sqrt{\pi }}\int _0^z e^{-t^2} dt.$

}\\
 &\\
Normal $(\mu,\sigma)$ with switching variance $\sigma^2 a$ w.p $p$\footnote{See comments and derivations in the appendix for switching both variance and mean as it can produce negative values for kappa.}. & $2-\frac{\log (2)}{\log \left(\frac{\sqrt{2} \left(\sqrt{\frac{a p}{p-1}+\sigma ^2}+p \left(-2 \sqrt{\frac{a p}{p-1}+\sigma ^2}+p \left(\sqrt{\frac{a p}{p-1}+\sigma ^2}-\sqrt{2 a \left(\frac{1}{p-1}+2\right)+4 \sigma ^2}+\sqrt{a+\sigma ^2}\right)+\sqrt{2 a \left(\frac{1}{p-1}+2\right)+4 \sigma ^2}\right)\right)}{p \sqrt{a+\sigma ^2}-(p-1) \sqrt{\frac{a p}{p-1}+\sigma ^2}}\right)}$\\
 &\\
 Lognormal $(\mu,\sigma)$ & $\approx 2-\frac{\log (2)}{\log \left(\frac{2 \text{ erf}\left(\frac{\sqrt{\log \left(\frac{1}{2} \left(e^{\sigma ^2}+1\right)\right)}}{2 \sqrt{2}}\right)}{\text{ erf}\left(\frac{\sigma }{2 \sqrt{2}}\right)}\right)}$. 
\end{tabular}
\end{mdframed}
\end{table*}

\begin{table}[h!]
\caption{Main results}	
\begin{mdframed}\label{tableParetoStudent}
\begin{tabular}{p{3cm}|p{4cm}}

  \textbf{Distribution} & $\kappa_n$ \\
  \hline
&\\
 Exponential/Gamma & Explicit  \\
 &\\
 Lognormal $(\mu,\sigma)$ & No explicit $\kappa_n$ but explicit lower and higher bounds (low or high $\sigma$ or $n$). Approximated with Pearson IV for $\sigma$ in between.\\
 & \\
 Pareto $(\alpha)$ (Constant) & Explicit for $\kappa_2$ (lower bound for all $\alpha$).\\
 & \\
 Student T$(\alpha)$ (slowly varying function)& Explicit for $\kappa_1$  , $\alpha =3$.\\

\end{tabular}
\end{mdframed}
\end{table}

\begin{table}
\caption{Comparing Pareto to Student T (Same tail exponent $\alpha$)}

\begin{tabular}{l lllll l}
 $\alpha$ &\text{Pareto } & \text{Pareto } & \text{Pareto } & \text{Student} &
   \text{Student} & \text{Student}\\
     &$\kappa_1$ & $\kappa_{1,30}$ & $\kappa_{1,100}$&$\kappa_1$ & $\kappa_{1,30}$ &
   $\kappa_{1,100}$ \\
 1.25 & 0.829& 0.787& 0.771& 0.792& 0.765& 0.756 \\
 1.5 & 0.724& 0.65& 0.631& 0.647& 0.609& 0.587\\
 1.75 & 0.65& 0.556& 0.53& 0.543& 0.483& 0.451 \\
 2. & 0.594& 0.484& 0.449& 0.465& 0.387& 0.352 \\
 2.25 & 0.551& 0.431& 0.388& 0.406& 0.316& 0.282 \\
 2.5 & 0.517& 0.386& 0.341& 0.359& 0.256& 0.227\\
 2.75 & 0.488& 0.356& 0.307& 0.321& 0.224& 0.189 \\
 3. & 0.465& 0.3246& 0.281& 0.29& 0.191& 0.159 \\
 3.25 &  0.445& 0.305& 0.258& 0.265& 0.167& 0.138 \\
 3.5 & 0.428& 0.284& 0.235& 0.243& 0.149& 0.121 \\
 3.75 & 0.413& 0.263& 0.222& 0.225& 0.13& 0.10 \\
 4. &  0.4& 0.2532& 0.211& 0.209& 0.126& 0.093 \\
\end{tabular}
\end{table}

 \begin{figure}[h!]
\includegraphics[width=\linewidth]{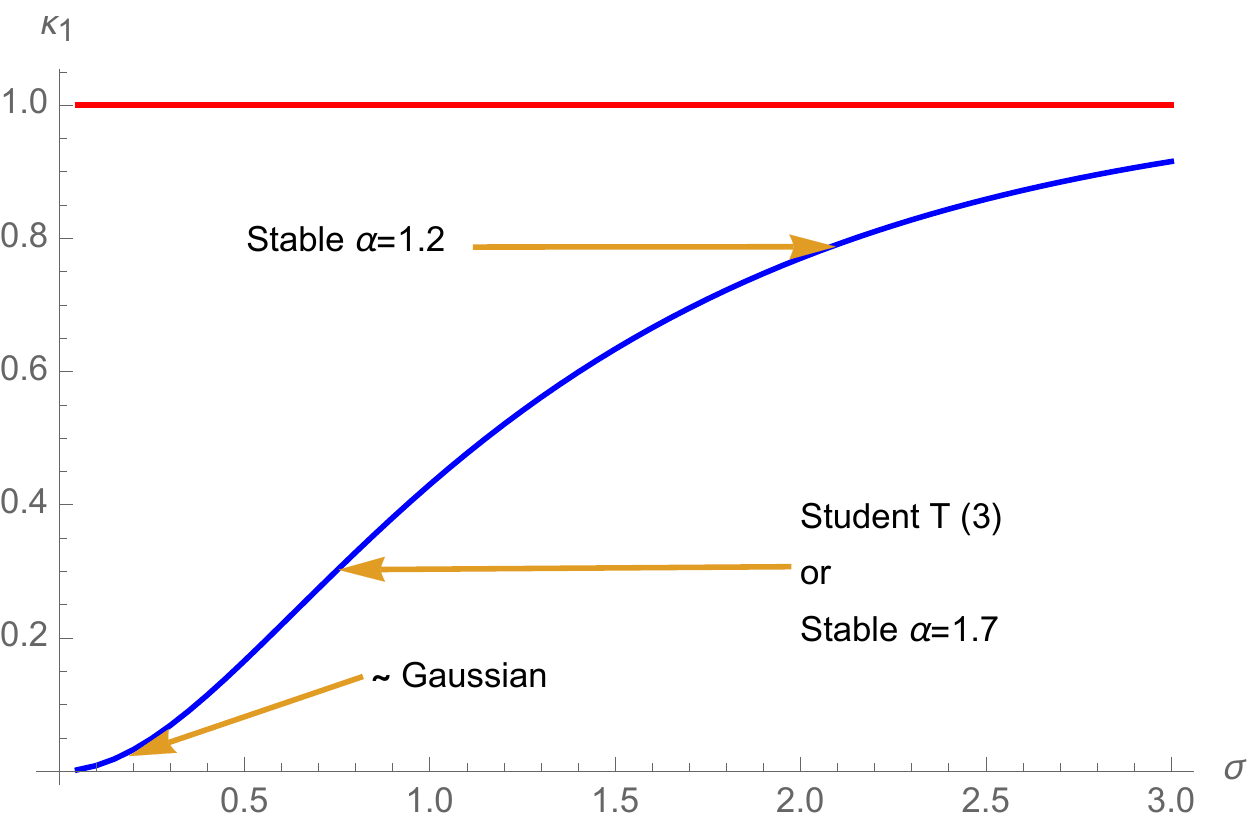}\label{lognormalgraph}
\caption{The lognormal distribution behaves like a Gaussian for low values of $\sigma$, but becomes rapidly equivalent to a power law. This illustrates why, operationally, the debate on whether the distribution of wealth was lognormal (Gibrat) or Pareto (Zipf) doesn't carry much operational significance.} 
\end{figure}

\begin{definition}[the $\kappa$ metric]
	Let $X_1, \dots, X_n$ be i.i.d. random variables with finite mean, that is $\mathbb{E}(X)< +\infty $. Let $S_n= X_1+X_2+\ldots +X_n$ be a partial sum.
	Let $\mathbb{M}(n)=\mathbb{E}(|S_n-\mathbb{E}(S_n)|)$ be the expected mean absolute deviation from the mean for $n$ summands. Define the "rate" of convergence for $n$ additional summands starting with $n_0$: 

\begin{equation*}
 \kappa_{n_0,n} = \min\left\{\kappa_{n_0,n}:\frac{\mathbb{M}(n)}{ \mathbb{M}(n_0)}= \left(\frac{n}{n_0}\right)^{\frac{1}{2-\kappa_{n_0,n}}} \,, n_0=1,2,...	\right \},
\end{equation*}
$n > n_0\geq 1$, hence
\begin{equation}
	\kappa(n_0,n)= 2-\frac{\log (n)-\log(n_0)}{\log \left(\frac{\mathbb{M}(n)}{\mathbb{M}(n_0)}\right)}.
\end{equation}

Further, for the baseline values $n=n_0+1$, we use the shorthand $\kappa_{n_0}$.
\end{definition}

We can also decompose $\kappa(n_0,n)$ in term of "local" intermediate ones similar to "local" interest rates, under the constraint.
\begin{equation}
	\kappa(n_0,n)= 2-\frac{\log (n)-\log (n_0)}{\sum _{i=0}^{n} \frac{\log (i+1)-\log (i)}{2-\kappa(i,i+1)}}.
\end{equation}

\subsubsection*{Use of Mean Deviation} Note that we use for measure of dispersion around the mean the mean absolute deviation, to stay in norm $L^1$ in the absence of finite variance --actually, even in the presence of finite variance, under power law regimes, distributions deliver an unstable and uninformative second moment. Mean deviation proves far more robust there. (Mean absolute deviation can be shown to be more "efficient" except in the narrow case of kurtosis equals 3 (the Gaussian), see a longer discussion in \cite{talebtechnicalincerto2018}; for other advantages, see \cite{pham2001mean}.)

\section{Stable Basin of Convergence as Benchmark}



\begin{definition}[the class $\mathfrak{P}$]
	The $\mathfrak{P}$ class of power laws (regular variation) is defined for r.v. $X$ as follows:
\begin{equation}
\mathfrak{P}=\{X:\mathbb{P}(X>x)\sim L(x)\,x^{-\alpha } \} \label{powerlaweq2}
\end{equation}%
where $\sim$ means that the limit of the ratio or rhs to lhs goes to 1 as $x \to \infty$. $L:\left[ x_{\min },+\infty \right) \rightarrow \left( 0,+\infty
\right) $ is a slowly varying function, defined as $\lim_{x\rightarrow
+\infty }\frac{L(kx)}{L(x)}=1$ for any $k>0$. The constant $\alpha >0$.

\end{definition}
Next we define the domain of attraction of the sum of identically distributed variables, in our case with identical parameters.

\begin{definition} (stable $\mathfrak{S}$ class)
A random variable $X$ follows a stable (or $\alpha$-stable) distribution,  symbolically $X\sim S(\tilde\alpha,\beta,\mu,\sigma)$, if its characteristic function $\chi(t)=\mathbb{E}(e^{itX})$ is of the form:
\begin{equation}
\chi(t)=\begin{cases}
e^{ \left(i \mu  t-\left| t \sigma \right| ^{\tilde{\alpha} } \left(1-i \beta  \tan \left(\frac{\pi  \tilde{\alpha} }{2}\right) \sgn(t)\right)\right)}& \tilde{\alpha} \ne 1 \\
\\
e^{i t \left(\frac{2 \beta  \sigma  \log (\sigma )}{\pi }+\mu \right)-\left| t \sigma \right|  \left(1+\frac{2 i \beta  \sgn(t) \log (\left| t \sigma \right| )}{\pi }\right)} & \tilde{\alpha} = 1\label{stable}
\end{cases},
\end{equation}
%
\end{definition}

Next, we define the corresponding stable $\tilde{\alpha}$:
\begin{equation}
\tilde{\alpha} \triangleq \begin{cases}  \alpha \,\mathbbm{1}_{\alpha<2} + 2 \,\mathbbm{1}_{\alpha\geq 2} & \text{if }$X$ \text{ is in }  \mathfrak{P}\\  2 &\mbox{otherwise}.\end{cases}
\end{equation}

 Further discussions of the class $\mathfrak{S}$ are as follows.
 
 \subsection{Equivalence for stable distributions}
For all $n_0$ and $n \geq 1$ in the Stable $\mathfrak{S}$ class with $\tilde\alpha \geq 1$:
$$\kappa_{(n_0,n)}=2-\tilde{\alpha},$$
simply from the property that
\begin{equation}
\mathbb{M}(n)=n^{\frac{1}{\alpha}} \mathbb{M}(1)	
\end{equation}

This, simply shows that $\kappa_{n_0,n}=0$ for the Gaussian.




 The problem of the preasymptotics for $n$ summands reduces to:
 \begin{itemize}
  \item What is the property of the distribution for $n_0=1$ (or starting from a standard, off-the shelf distribution)?	
 \item What is the property of the distribution for $n_0$ summands?	
\item How does $\kappa_n \rightarrow 2-\tilde{\alpha}$ and at what rate? 
 \end{itemize}  

\subsection{Practical significance for sample sufficiency}
\begin{mdframed}[backgroundcolor=red!20]
\textbf{Confidence intervals}: As a simple heuristic, the higher $\kappa$, the more disproportionally insufficient the confidence interval. Any value of $\kappa$ above .15 effectively indicates a high degree of unreliability of the "normal approximation". One can immediately doubt the results of numerous research papers in fat-tailed domains.
\end{mdframed}

Computations of the sort done Table \ref{tableParetoStudent} for instance allows us to compare various distributions under various parametriazation. (comparing various Pareto distributions to symmetric Student T and, of course the Gaussian which has a flat kappa of 0) 
 
As we mentioned in the introduction, required sample size for statistical inference is driven by $n$, the number of summands. Yet the law of large numbers is often invoked in erroneous conditions; we need a rigorous sample size metric. 

Many papers, when discussing financial matters, say \cite{gabaix2008power} use finite variance as a binary classification for  fat tailedness: power laws with a tail exponent greater than $2$ are therefore classified as part of the "Gaussian basin", hence allowing the use of variance and other such metrics for financial applications.  A much more natural boundary is finiteness of expectation for financial applications \cite{taleb2009finiteness}. Our metric can thus be useful as follows:

Let $X_{g,1},X_{g,2},\ldots, X_{g,n_g}$ be a sequence of Gaussian variables with mean $\mu$ and scale $\sigma$. Let $X_{\nu,1},X_{nu,2},\ldots, X_{nu,n_\nu}$ be a sequence of some other variables scaled to be of the same $\mathbb{M}(1)$, namely $\mathbb{M}^\nu(1)=\mathbb{M}^g(1)=\sqrt{\frac{2}{\pi}}\sigma$. We would be looking for values of $n_\nu$ corresponding to a given $n_g$.
\begin{mdframed}[backgroundcolor=red!20]
$\kappa_n$ is indicative of both the rate of convergence under the law of large number, and for $\kappa_n \rightarrow 0$, for rate of convergence of summands to the Gaussian under the central limit, as illustrated in Figure \ref{convergencetoCLT}.
\end{mdframed}
\begin{multline*}
n_{\min}=
\inf\left\{n_\nu:\mathbb{E}\left(\left|\sum_{i=1}^{n_\nu} \frac{X_{\nu,i}-m_p}{n_\nu}\right|\right)\right. \\ \left.\leq \mathbb{E}\left(\left|\sum_{i=1}^{n_g} \frac{X_{g,i}-m_g}{n_g}\right|\right),n_\nu>0\right\}
\label{md}
\end{multline*}
which can be computed using $\kappa_n=0$ for the Gaussian and backing our from $\kappa_n$ for the target distribution with the simple approximation:
\begin{equation}
n_\nu= n_g^{-\frac{1}{\kappa _{1,n_g}-1}}\approx	 n_g^{-\frac{1}{\kappa _{1}-1}}\,,  n_g >1 \label{simpleequiv}
\end{equation}
The approximation is owed to the slowness of convergence. So for example, a Student T with 3 degrees of freedom ($\alpha=3$) requires 120 observations to get the same drop in variance from averaging (hence confidence level) as the Gaussian with 30, that is 4 times as much.  The one-tailed Pareto with the same tail exponent $\alpha=3$ requires 543 observations to match a Gaussian sample of 30,  4.5 times more than the Student, which shows 1)  finiteness of variance is not an indication of fat tailedness (in our statistical sense), 2) neither are tail exponents good indicators 3) how the symmetric Student and the Pareto distribution are not equivalent because of the "bell-shapedness" of the Student (from the slowly moving function) that dampens variations in the center of the distribution.

We can also elicit quite counterintuitive results. From Eq. \ref{simpleequiv}, the "Pareto 80/20" in the popular mind, which maps to a tail exponent around $\alpha  \approx 1.14$, requires $>10^9$ more observations than the Gaussian.
\section{Technical Consequences}

\subsection{Some oddities with asymmetric distributions}

The stable distribution, when skewed, has the same $\kappa$ index as a symmetric one (in other words,  $\kappa$  is invariant to the $\beta$ parameter in Eq. \ref{stable}, which  conserves under summation). But a one-tailed simple Pareto distribution is fatter tailed (for our purpose here) than an equivalent symmetric one. 

This is relevant because the stable is never really observed in practice and used as some limiting mathematical object, while the Pareto is more commonly seen. The point is not well grasped in the literature.  Consider the following use of the substitution of a stable for a Pareto. In Uchaikin and Zolotarev \cite{uchaikin1999chance}: 
\begin{quote}Mandelbrot called attention to the fact that the use of the extremal stable distributions (corresponding to $\beta = 1$) to describe empirical principles was preferable to the use of the Zipf-Pareto distributions for a number of reasons. It can be seen from many publications, both theoretical and applied, that Mandelbrot's ideas receive more and more wide recognition of experts. In this way, the hope arises to confirm empirically established principles in the framework of mathematical models and, at the same time, to clear up the mechanism of the formation of these principles.	
\end{quote}
These are not the same animals, even for large number of summands. 

\subsection{Rate of convergence of a student T  distribution to the Gaussian Basin}
We show in the appendix --thanks to the explicit derivation of $\kappa$ for the sum of students with $\alpha=3$, the "cubic" commonly noticed in finance  --that the rate of convergence of $\kappa$ to $0$ under summation is slow. The semi-closed form for the density of an n-summed cubic Student allows to complement the result in Bouchaud and Potters \cite{bouchaud2003theory} (see also \cite{sornette2004critical}, which is as follows. Their approach is to separate the "Gaussian zone" where the density is approximated by that of a Gaussian, and a "power law zone" in the tails which retains the original distribution with power law decline. The "crossover" between the two moves right and left of the center at a rate of $\sqrt{n \log (n)}$ standard deviations) which is excruciatingly slow. Indeed, one can note that more summands fall at the center of the distribution, and fewer outside of it, hence the speed of convergence according to the central limit theorem will differ according to whether the density concerns the center or the tails.

Further investigations would concern the convergence of the Pareto to a Levy-Stable, which so far we only got numerically.

\subsection{The lognormal is neither thin nor fat tailed}
Naively, as we can see in Figure \ref{lognormalgraph}, at low values of the parameter $\sigma$, the lognormal  behaves like a Gaussian, and, at high $\sigma$, it appears to have the behavior of  a Cauchy of sorts (a one-tailed Cauchy, rather a stable distribution with $\alpha=1$, $\beta=1$), as $\kappa$ gets closer and closer to $1$. This gives us an idea about some aspects of the debates as to whether some variable is Pareto or lognormally distributed, such as, say, the debates about wealth \cite{mandelbrot1960pareto}, \cite{dagum1980inequality}, \cite{dagum1983income}. Indeed, such debates can be irrelevant to the real world. As P. Cirillo \cite{cirillo2013your} observed,  many cases of Paretianity are effectively lognormal situations with high variance; the practical statistical consequences, however, are smaller than imagined.



%

\subsection{Can kappa be negative?}
Just as kurtosis for a mixed Gaussian (i.e., with stochastic mean, rather than stochastic volatility) can dip below $3$ (or become "negative" when one uses the convention of measuring Kurtosis as excess over the Gaussian by adding $3$ to the measure), the kappa metric can become negative when kurtosis is "negative". These situations require bimodality (i.e., a switching process between means under fixed variance, with modes far apart in terms of standard deviation). They do not appear to occur with unimodal distributions.

Details and derivations are presented in the appendix.
 
\section{Conclusion and Consequences}

To summarize, while the limit theorems (the law of large numbers and the central limit) are concerned with the behavior as $n\to +\infty$, we are interested in finite and exact $n$ both small and large (and its statistical and risk implications).

We may draw a few operational consequences:
\begin{figure}[h!]
\includegraphics[width=\linewidth]{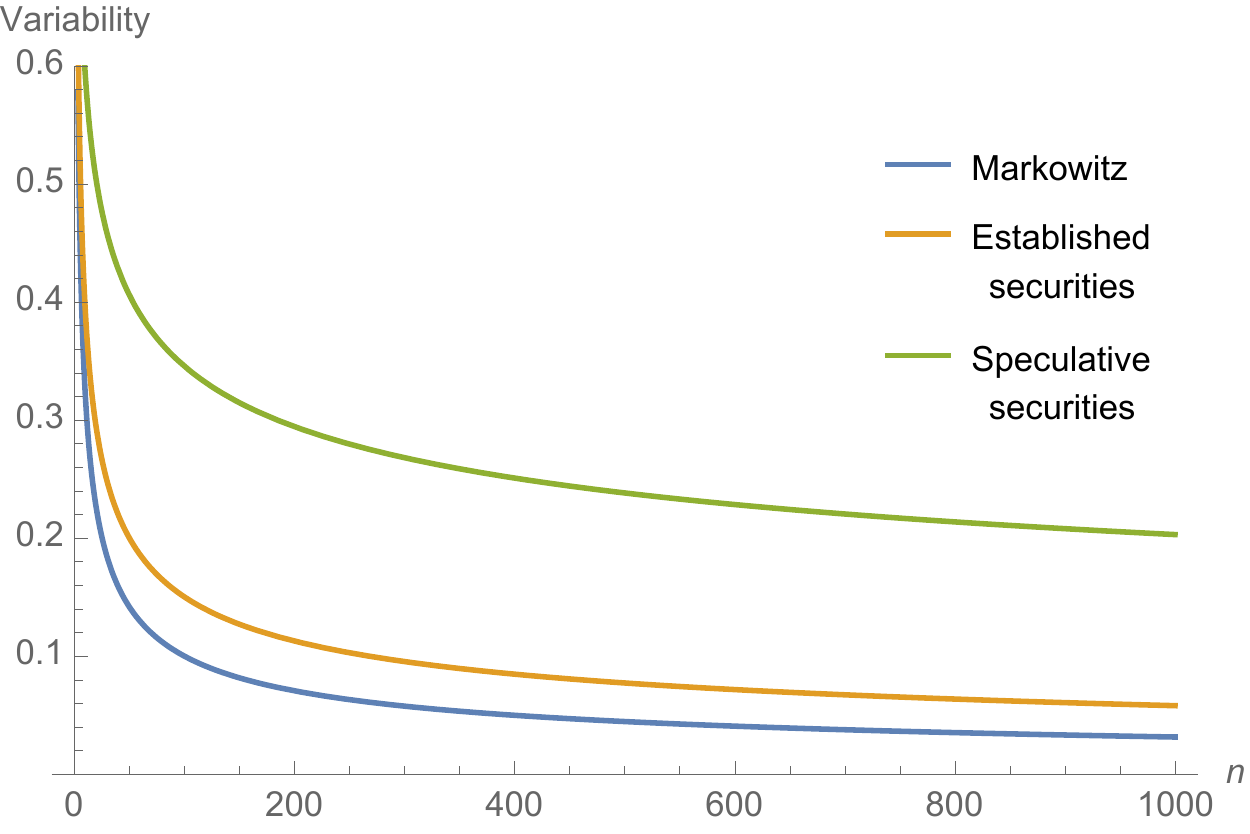}\label{oneovern}
	\caption{In short, why the 1/n heuristic works in portfolio theory (and similar decision problems): it takes many, many more securities to get the same risk reduction as via portfolio allocation according to Markowitz. We assume to simplify that the securities are independent, which they are not, something that compounds the effect. }
\end{figure}
\subsection{Portfolio pseudo-stabilization}

 Our method can also naturally and immediately apply to portfolio construction and the effect of diversification since adding a security to a portfolio has the same "stabilizing" effect as adding an additional observation for the purpose of statistical significance. "How much data do you need?" translates into "How many securities do you need?". Clearly, the Markowicz allocation method in modern finance\cite{markowitz1952portfolio} (which seems to not be used by Markowitz himself for his own portfolio \cite{neth2015heuristics}) applies only for $\kappa$ near $0$; people use convex heuristics, otherwise they will underestimate tail risks and "blow up" the way the famed portfolio-theory oriented hedge fund Long Term Management did in 1998 \cite{taleb2018skin} \cite{thorp1969optimal}.)

We mentioned earlier that a Pareto distribution close to the "80/20" requires up to $10^{9}$ more observations than a Gaussian; consider that the risk of a portfolio under such a distribution would be underestimated by at least $8$ orders of magnitudes if one uses modern portfolio criteria. Following such a reasoning, one simply needs broader portfolios. 

It has also been noted that there is practically no financial security that is not fatter tailed than the Gaussian, from the simple criterion of kurtosis \cite{taleb2009errors}, meaning Markowitz portfolio allocation is \textit{never} the best solution. It happens that agents wisely apply a noisy approximation to the $\frac{1}{n}$ heuristic which has been classified as one of those biases by behavioral scientists but has in fact been debunked as false (a false bias is one in which, while the observed phenomenon is there, it does not constitute a "bias" in the bad sense of the word; rather  it is the researcher who is mistaken owing to using the wrong tools instead of the decision-maker).  This tendency to "overdiversify"  has been deemed a departure from optimal investment behavior by  Benartzi and Thaler \cite{benartzi2001naive}, explained in \cite{benartzi2007heuristics} "when faced with $n$ options, divide assets evenly across the options. We have dubbed this heuristic the "$1/n$ rule."" However,  broadening one's diversification is effectively as least as optimal as standard allocation(see critique by Windcliff and Boyle \cite{windcliff20041} and \cite{demiguel2007optimal}). In short, an equally weighted portfolio outperforms the SP500 across a broad range range of metrics. But even the latter two papers didn't conceive of the full effect and properties of fat tails, which we can see here with some precision. Fig. \ref{oneovern} shows the effect for securities compared to Markowitz.

This false bias is one in many examples of policy makers "nudging" people into the wrong rationality \cite{taleb2018skin} and driving them to increase their portfolio risk many folds.

A few more comments on financial portfolio risks. The SP500 has a $\kappa$ of around .2, but one needs to take into account that it is itself a basket of $n=500$ securities, albeit unweighted and consisting of correlated members, overweighing stable stocks. Single stocks have kappas between $.3$ and $.7$, meaning a policy of "overdiversification" is a must.

Likewise the metric gives us some guidance in the treatment of data for forecasting, by establishing sample sufficiency, to state such matters as how many years of data do we need before stating whether climate conditions "have changed", see \cite{makridakis2009decision}.

\subsection{Other aspects of statistical inference}
So far we considered only univariate distributions. For higher dimensions, a potential area of investigation is an equivalent approach to the multivariate distribution of fat tailed variables, the sampling of which is not captured by the Marchenko-Pastur (or Wishhart) distributions. As in our situation, adding variables doesn't easily remove noise from random matrices.
\subsection{Final comment}
As we said earlier, "statistics is never standard"; however there are heuristics methods to figure out where and by how much we depart from the standard. 
\onecolumn
\appendix

 \subsection{Cubic Student T (Gaussian Basin)}
The Student T with 3 degrees of freedom is of special interest in the literature owing to its prevalence in finance \cite{gabaix2008power}. It is often mistakenly approximated to be Gaussian owing to the finiteness of its variance. Asymptotically, we end up with a Gaussian, but this doesn't tell us anything about the rate of convergence. Mandelbrot and Taleb \cite{mandelbrot2010random} remarks that the cubic acts more like a powerlaw in the distribution of the extremes, which we will elaborate here thanks to an explicit PDF for the sum.

Let $X$ be a random variable distributed with density $p(x)$:
\begin{equation}
p(x)=\frac{6 \sqrt{3}}{\pi  \left(x^2+3\right)^2} \; , x \in (-\infty,\infty) \label{pdfcubic}
	\end{equation}
\begin{proposition}

	Let $Y$ be a sum of $X_1, \dots, X_n$, $n$ identical copies of $X$. Let $\mathbb{M}(n)$ be the mean absolute deviation from the mean for $n$ summands. The "rate" of convergence $\kappa_{1,n} =\left\{{\kappa:\frac{\mathbb{M}(n)}{\mathbb{M}(1)}=n^{\frac{1}{2-\kappa }}}\right\}$ is:
\begin{equation}
 \kappa_{1,n}=2-\frac{\log (n)}{\log \left(e^n n^{-n} \Gamma (n+1,n)-1\right)}
\end{equation}
\end{proposition}
where $\Gamma(.,.)$ is the incomplete gamma function $\Gamma (a,z)=\int _z^{\infty }d t t^{a-1} e^{-t}$.
 
Since the mean deviation $\mathbb{M}(n)$:

\begin{equation}
\mathbb{M}(n)=\begin{cases}
	\frac{2 \sqrt{3}}{\pi } & \text{for } \, n=1\\
	 \frac{2 \sqrt{3} }{\pi }\left(e^n n^{-n} \Gamma (n+1,n)-1\right)& \text{for }\, n>1\\
\end{cases}	\label{MDcubic}
\end{equation}

The derivations are as follows. For the pdf and the MAD we followed different routes.

We have the characteristic function for $n$ summands:
$$\varphi(\omega) = (1+\sqrt{3}|\omega|)^n\,e^{-n\sqrt{3}\,|\omega|}$$
The pdf of $Y$ is given by:
\begin{dmath*}
p(y)=\frac{1}{\pi}\int_{0}^{\infty}(1+\sqrt{3}\,\omega)^n\,e^{-n\sqrt{3}\,\omega}\cos(\omega y)\,d\omega
\end{dmath*}
After arduous integration we get the result in \ref{MDcubic}. 
	Further, since the following result does not appear to be found in the literature, we have a side useful result: the PDF of $Y$ can be written as
	
	\begin{equation}
p(y) =\frac{e^{n-\frac{i y}{\sqrt{3}}} \left(e^{\frac{2 i y}{\sqrt{3}}} E_{-n}\left(n+\frac{i y}{\sqrt{3}}\right)+E_{-n}\left(n-\frac{i y}{\sqrt{3}}\right)\right)}{2 \sqrt{3} \pi }	\label{pdfcubicsum}
\end{equation}
where $E_{(.)}(.)$ is the exponential integral $E_n z=\int _1^{\infty }\frac{ e^{t (-z)}}{t^n}dt$.

 Note the following  identities (from the updating of Abramowitz and Stegun) \cite{NIST:DLMF}
 $$n^{-n-1} \Gamma (n+1,n)= E_{-n}(n)=e^{-n}\frac{(n-1)!}{n^n} \sum _{m=0}^n \frac{n^m}{m!}$$

As to the asymptotics, we have the following result (proposed by Michail Loulakis):
Reexpressing Eq. \ref{MDcubic}:

$$\mathbb{M}(n)=\frac{2 \sqrt{3} n! }{\pi  n^n}\sum _{m=0}^{n-1} \frac{n^m}{m!}$$
Further, $$e^{-n}\sum _{m=0}^{n-1} \frac{n^m}{m!}= \frac{1}{2} +O\left(\frac{1}{\sqrt{n}}\right)$$
(From the behavior of the sum of Poisson variables as they converge to a Gaussian by the central limit theorem: $e^{-n} \sum _{m=0}^{n-1} \frac{n^m}{m!} = \mathbb{P}(X_n < n)$ where $X_n$ is a Poisson random variable with parameter $n$.  Since the sum of $n$ independent Poisson random variables with parameter $1$ is Poisson with parameter $n$, the Central Limit Theorem says the probability distribution of $Z_n = (X_n - n)/\sqrt{n}$ approaches a standard normal distribution.  Thus $\mathbb{P}(X_n < n) = \mathbb{P}(Z_n < 0) \to 1/2$ as $n \to \infty$.\footnote{Robert Israel on Math Stack Exchange}     For another approach, see \cite{newman2012problem} for proof that $ 1 + \frac{n}{1!} + \frac{n^2}{2!} + \dots + \frac{n^{n-1}}{(n-1)!} \sim
\frac{e^n}{2}$.)

Using the property that $\underset{n\to \infty }{\text{lim}}\frac{n! \exp (n)}{n^n \sqrt{n}}=\sqrt{2 \pi }$, we get the following exact asymptotics:

$$\lim_{n\to \infty} \log(n) \kappa_{1,n}=\frac{\pi^2}{4}$$
thus $\kappa$ goes to 0 (i.e, the average becomes Gaussian) at speed $\frac{1}{\log(n)}$, which is excruciatingly slow. In other words, even with $10^6$ summands, the behavior cannot be summarized as that of a Gaussian, an intuition often expressed by B. Mandelbrot \cite{mandelbrot2010random}.

\subsection{Lognormal Sums}
From the behavior of its cumulants for $n$ summands, we can observe that a sum behaves likes a Gaussian when $\sigma$ is low, and as a lognormal when $\sigma$ is high --and in both cases we know explicitly $\kappa_n$. 

The lognormal (parametrized with $\mu$ and $\sigma$) doesn't have an explicit characteristic function. But we can get cumulants $K_i$ of all orders $i$ by recursion and for our case of summed identical copies of r.v. $X_i$, $K_i^n=K_i(\sum_n X_i)=nK_i(X_1)$. 

Cumulants:
$$\begin{array}{l}
K_1^n = n e^{\mu +\frac{\sigma ^2}{2}} \\
K_2^n= n \left(e^{\sigma ^2}-1\right) e^{2 \mu +\sigma ^2} \\
K_3^n= n \left(e^{\sigma ^2}-1\right)^2 \left(e^{\sigma ^2}+2\right) e^{3 \mu +\frac{3 \sigma
   ^2}{2}} \\
K_4^n= \ldots 
\end{array}
$$

Which allow us to compute:  $Skewness= \frac{\sqrt{e^{\sigma ^2}-1} \left(e^{\sigma ^2}+2\right) e^{\frac{1}{2} \left(2 \mu +\sigma ^2\right)-\mu -\frac{\sigma ^2}{2}}}{\sqrt{n}}$
and 
$Kurtosis= 3+\frac{e^{2 \sigma ^2} \left(e^{\sigma ^2} \left(e^{\sigma ^2}+2\right)+3\right)-6}{n}$

We can immediately prove from the cumulants/moments that:
$$ \lim_{n \to +\infty }\kappa_{1,n}=0, 
 \lim _{\sigma \to 0 }\kappa_{1,n}=0$$



and our bound on $\kappa$ becomes explicit:

Let $\kappa^*_{1,n}$ be the situation under which the sums of lognormal conserve the lognormal density, with the same first two moments. We have
$$0\leq \kappa^*_{1,n}\leq 1,$$


$$\kappa^*_{1,n}=2-\frac{\log (n)}{\log \left(\frac{n \text{erf}\left(\frac{\sqrt{\log \left(\frac{n+e^{\sigma
   ^2}-1}{n}\right)}}{2 \sqrt{2}}\right)}{\text{erf}\left(\frac{\sigma }{2
   \sqrt{2}}\right)}\right)}$$
   
\subsubsection{Heuristic attempt} Among other heuristic approaches, we can see in two steps how 1) under high values of $\sigma$, $\kappa_{1,n} \to \kappa^*_{1,n}$, since the law of large numbers slows down, and 2) $\kappa^*_{1,n}\overset{\sigma \to \infty} {\to}  1$. 
%


\subsubsection{Loulakis' Proof} Proving the upper bound, that for high variance $\kappa_{1,n}$ approaches $1$ has been shown formally my Michail Loulakis\footnote{Review of this paper; Loulakis proposed a formal proof in place of the heuristic derivation.} which we summarize as follows.
We start with the identify 
	$\mathbb{E}\left(|X-m|\right)= 2 \int_m^\infty (x-m) f(x) dx=2 \int_m^\infty \bar F_X(t) dt \label{mdidentity},$
where $f(.)$ is the density, $m$ is the mean, and $\bar F_X(.)$ is the survival function.
Further, $\mathbb{M}(n)=2 \int_{n m}^\infty \bar F(x) dx .$ Assume $\mu=\frac{1}{2} \sigma^2$, or $X=\exp \left(\sigma  Z-\frac{\sigma ^2}{2}\right)$ where Z is a standard normal variate. Let $S_n$ be the sum $X_1 +\ldots +X_n$; we get $\mathbb{M}(n)=2 \int_{n}^\infty \mathbb{P}(S_n>t) dt$. 
Using the property of subexponentiality (\cite{pitman1980subexponential}), 
$\mathbb{P}(S_n>t) \geq \mathbb{P}(\max_{0<i\leq n} (X_i) > t) \geq n \mathbb{P}(X_1>t) -\binom{n}{2}\mathbb{P}\left(X_1>t\right)^2$. Now $\mathbb{P}\left(X_1>t\right) \overset{\sigma \to \infty} {\to} 1$ and the second term to $0$ (using H\"older's inequality). 

Skipping steps, we get $\underset{\sigma \to \infty} {\text{lim inf}} \frac{\mathbb{M}(n)}{\mathbb{M}(1)}\geq n$, while at the same time we need to satisfy the bound $\frac{\mathbb{M}(n)}{\mathbb{M}(1)} \leq n$. So for $\sigma \to \infty$ ,$\frac{\mathbb{M}(n)}{\mathbb{M}(1)}=n$, hence $\kappa_{1,n}\overset{\sigma \to \infty} {\to}  1$.


\subsubsection{Pearson Family approach for computation} 
For computational purposes, for the $\sigma$ parameter not too large (below $\approx .3$, we can use the Pearson family for computational convenience --although the lognormal does not belong to the Pearson class (the normal does, but we are close enough for computation). Intuitively, at low $sigma$, the first four moments can be sufficient because of the absence of large deviations; not at higher $sigma$ for which conserving the lognormal would be the right method.

The use of Pearson class is practiced in some fields such as information/communication theory, where there is a rich literature: for summation of lognormal variates see Nie and Chen, \cite{nie2007lognormal}, and for Pearson IV, \cite{chen2008lognormal}, \cite{di2009further}.

The Pearson family is defined for an appropriately scaled density $f$ satisfying the following differential equation.
\begin{equation}
f'(x)=-\frac{(a_0+a_1 x)}{b_0+b_1 x+b_2 x^2}f (x)	
\end{equation}

 We note that our parametrization of $a_0$, $b_2$, etc. determine the distribution within the Pearson class --which appears to be the Pearson IV.  Finally we get an expression of mean deviation as a function of $n$, $\sigma$, and $\mu$.

Let $m$ be the mean. Diaconis et al \cite{diaconis1991closed} from an old trick by De Moivre, Suzuki \cite{suzuki1965consistent} show that we can get explicit mean absolute deviation.
Using, again, the identity $\mathbb{E}(|X-m|)=2 \int _{m }^\infty (x-m) f(x)\mathrm{d}x$ and integrating by parts,
\begin{equation}
\mathbb{E}(|X-m|)=\frac{2 \left(b_0+b_1 m +b_2 m^2\right)}{a_1-2 b_2}f(m)	
\end{equation}

We use cumulants of the n-summed lognormal to match the parameters. Setting $a_1=1$, and $m=\frac{b_1-a_0}{1-2 b_2}$, we get
$$\left\{
\begin{array}{l}
a_0= \frac{e^{\mu +\frac{\sigma ^2}{2}} \left(-12 n^2+(3-10 n) e^{4 \sigma ^2}+6 (n-1)
   e^{\sigma ^2}+12 (n-1) e^{2 \sigma ^2}-(8 n+1) e^{3 \sigma ^2}+3 e^{5 \sigma ^2}+e^{6 \sigma
   ^2}+12\right)}{2 \left(6 (n-1)+e^{2 \sigma ^2} \left(e^{\sigma ^2} \left(5 e^{\sigma
   ^2}+4\right)-3\right)\right)} \\
b_2= \frac{e^{2 \sigma ^2} \left(e^{\sigma ^2}-1\right) \left(2 e^{\sigma ^2}+3\right)}{2
   \left(6 (n-1)+e^{2 \sigma ^2} \left(e^{\sigma ^2} \left(5 e^{\sigma
   ^2}+4\right)-3\right)\right)} \\
b_1= \frac{\left(e^{\sigma ^2}-1\right) e^{\mu +\frac{\sigma ^2}{2}} \left(e^{\sigma ^2}
   \left(e^{\sigma ^2} \left(e^{\sigma ^2} \left(-4 n+e^{\sigma ^2} \left(e^{\sigma
   ^2}+4\right)+7\right)-6 n+6\right)+6 (n-1)\right)+12 (n-1)\right)}{2 \left(6 (n-1)+e^{2 \sigma
   ^2} \left(e^{\sigma ^2} \left(5 e^{\sigma ^2}+4\right)-3\right)\right)} \\
b_0= -\frac{n \left(e^{\sigma ^2}-1\right) e^{2 \left(\mu +\sigma ^2\right)}
   \left(e^{\sigma ^2} \left(-2 (n-1) e^{\sigma ^2}-3 n+e^{3 \sigma ^2}+3\right)+6 (n-1)\right)}{2
   \left(6 (n-1)+e^{2 \sigma ^2} \left(e^{\sigma ^2} \left(5 e^{\sigma
   ^2}+4\right)-3\right)\right)} \\
\end{array}
\right.$$
\bigskip
\subsubsection{Polynomial expansions}
Other methods, such as Gram-Charlier expansions, such as
Schleher  \cite{schleher1977generalized}, Beaulieu,\cite{beaulieu1995estimating}, proved less helpful to obtain $\kappa_n$. At high values of $\sigma$, the approximations become unstable as we include higher order Lhermite polynomials. See review in Dufresne  \cite{dufresne2008sums} and \cite{dufresne2004log}.

%


\subsection{Exponential}
The exponential is the "entry level" fat tails, just at the border.
$$\begin{array}{cc}
 f(x)= & 
\begin{array}{cc}
 \lambda  e^{-\lambda  x}, & x\geq 0 .
 \end{array}
 \\
\end{array}$$
By convolution the sum $Z=X_1,X_2,\ldots X_n$ we get,  by recursion, since $f(y)=\int_0^y f(x) f(y-x) \, dx=\lambda ^2 y e^{-\lambda  y}$:
\begin{equation}
	f_n(z)=\frac{\lambda ^n z^{n-1} e^{-\lambda  z}}{(n-1)!}
\end{equation}
which is the gamma distribution; we get the mean deviation for $n$ summands:
\begin{equation}
\mathbb{M}(n)=\frac{2 e^{-n} n^n}{\lambda  \Gamma (n)},
\end{equation}
hence:
\begin{equation}
	\kappa _{1,n}
	=2-\frac{\log (n)}{n \log (n)-n-\log (\Gamma (n))+1}
\end{equation}
	
	We can see the asymptotic behavior is equally slow (similar to the student) although the exponential distribution is sitting at the cusp of subexponentiality:
	$$\underset{n\to \infty }{\text{lim}}\log (n) \kappa_{1,n}=4-2 \log (2 \pi )$$
\newpage
\subsection{Negative kappa}
Consider the simple case of a Gaussian with switching means and variance: with probability $\frac{1}{2}, X \sim \mathcal{N}(\mu_1,\sigma_1)$ and with probability $\frac{1}{2}, X \sim \mathcal{N}(\mu_2,\sigma_2)$. The kurtosis will be 

\begin{equation}
	Kurtosis=3-\frac{2 \left(\left(\mu _1-\mu _2\right){}^4-6 \left(\sigma _1^2-\sigma
   _2^2\right)^2\right)}{\left(\left(\mu _1-\mu _2\right)^2+2 \left(\sigma _1^2+\sigma
   _2^2\right)\right)^2}
\end{equation}

As we see the kurtosis is a function of $d=\mu_1-\mu_2$. For situations where $\sigma_1=\sigma_2$, $\mu_1 \neq \mu_2$ , the kurtosis will be below that of the regular Gaussian, and our measure will naturally be negative. In fact for the kurtosis to remain above 3, $$ |d| \leq\sqrt[4]{6} \sqrt{\max(\sigma_1,\sigma_2)^2-\min(\sigma_1,\sigma_2)^2},$$ the stochasticity of the mean offsets the stochasticity of volatility.

These situations with thinner tails than the Gaussian are encountered with bimodal situations where $\mu_1$ and $\mu_2$ are separated; the effect becomes acute when they are separated by several standard deviations.
Let d$=\mu_1-\mu_2$ and $\sigma=\sigma_1=\sigma_2$ (to achieve minimum kurtosis),
\begin{equation}
\kappa_1=\frac{\log (4)}{\log (\pi )-2 \log \left(\frac{\sqrt{\pi } d e^{\frac{d^2}{4 \sigma ^2}} \text{erf}\left(\frac{d}{2 \sigma }\right)+2 \sqrt{\sigma ^2} e^{\frac{d^2}{4 \sigma ^2}}+2 \sigma }{d e^{\frac{d^2}{4 \sigma ^2}} \text{erf}\left(\frac{d}{2 \sqrt{2} \sigma }\right)+2 \sqrt{\frac{2}{\pi }} \sigma  e^{\frac{d^2}{8 \sigma ^2}}}\right)}+2
\end{equation}
which we see is negative for wide values of $\mu_1-\mu_2$.

\twocolumn
\bibliographystyle{IEEEtran}
\bibliography{/Users/nntaleb/Dropbox/Central-bibliography}

\end{document}